%
\def\@{{\char'100}}

\long\def\abstract#1{\bigskip{\advance\leftskip by 2true cm
\advance\rightskip by 2true cm\eightpoint\centerline{\bf
Abstract}\everymath{\scriptstyle}\vskip10pt\vbox{#1}}\bigskip}
\long\def\resume#1{{\advance\leftskip by 2true cm
\advance\rightskip by 2true cm\eightpoint\centerline{\bf
R\'esum\'e}\everymath{\scriptstyle}\vskip10pt \vbox{#1}}}

\def\references{\bigbreak\centerline{\sc
References}\medskip\nobreak\bgroup
\def\ref##1&{\leavevmode\hangindent 15pt
\hbox to 15pt{\hss\bf[##1]\ }\ignorespaces}
\parindent=0pt
\everypar={\ref}\par}
\def\endreferences{\egroup}
\long\def\authoraddr#1{\medskip{\baselineskip9pt\let\\=\cr
\halign{\line{\hfil{\Addressfont##}\hfil}\crcr#1\crcr}}}
\def\Subtitle#1{\medbreak\noindent{\Subtitlefont#1.} }
%
%
\newif\ifrunningheads
\runningheadstrue
\immediate\write16{- Page headers}
\headline={\ifrunningheads\ifnum\pageno=1\hfil\else\ifodd\pageno\rightheadline
\else\leftheadline\fi\fi\else\hfil\fi}
\def\rightheadline{\sc\hfil\RightHeadText\hfil}
\def\leftheadline{\sc\hfil\LeftHeadText\hfil}

\hyphenation{Harnad Neumann}
%
%
\immediate\write16{- Fonts "Small Caps" and "EulerFraktur"}  
%
%
%

\let\sc=\tensmc
%
%
\font\teneuf=eufm10  \font\seveneuf=eufm7 \font\fiveeuf=eufm5
\newfam\euffam \def\gr{\fam\euffam\teneuf}

\textfont\euffam=\teneuf \scriptfont\euffam=\seveneuf 
\scriptscriptfont\euffam=\fiveeuf
%
\edef\smatrix[#1\&#2\\#3\&#4]{\left({#1 \atop #3}\, {#2 \atop #4}\right)}

\def \smaller {\eightpoint}

\def \mt {\mapsto}
\def \ra {\rightarrow}

\def \a {\alpha}
\def \b {\beta}

\def \g {\gamma}
\def \G {\Gamma}

\def \l {\lambda}

\def \s {\sigma}
\def \S {\Sigma}
\def \th {\vartheta}
\def \t {\tau}
\def \o {\omega}

\def \ss {\subset}

\def \mod{{\rm mod\,}}

\def\nchi{\hbox{\raise 2.5pt\hbox{$\chi$}}}
%
%

\def\grG{{\gr G}}	
\def\grH{{\gr H}}

\def\nchi{\hbox{\raise 2.5pt\hbox{$\chi$}}}
%
%

%
%
		\def\bfA{{\bf A}}
		
		\def\bfC{{\bf C}}

		\def\bfI{{\bf I}}

		\def\bfQ{{\bf Q}}
		
		\def\bfS{{\bf S}}

		\def\bfZ{{\bf Z}}

%
%
\def\authorfont{\sc}
\font\eightrm=cmr8
\font\eightbf=cmbx8
\font\eightit=cmti8
\font\eightsl=cmsl8

\def\eightpoint{\let\rm=\eightrm \let\bf=\eightbf \let\it=\eightit
\let\sl=\eightsl \baselineskip = 9.5pt minus .75pt  \rm}

\font\titlefont=cmbx10 scaled\magstep2
\font\sectionfont=cmbx10
\font\Subtitlefont=cmbxsl10
\font\Addressfont=cmsl8
%
%
\def\Proclaim#1:#2\par{\smallbreak\noindent{\sc #1:\ }
{\sl #2}\par\smallbreak}
\def\Demo#1:#2\par{\smallbreak\noindent{\sl #1:\ }
{\rm #2}\par\smallbreak}
%
%
\immediate\write16{- Section headings}
\newcount\secount
\secount=0
\newcount\eqcount
\outer\def\section#1.#2\par{\global\eqcount=0\bigbreak
\ifcat#10
 \secount=#1\noindent{\sectionfont#1. #2}
\else
 \advance\secount by 1\noindent{\sectionfont\number\secount. #2}
\fi\par\nobreak\medskip} 
%
%
\immediate\write16{- Automatic numbering} 
\catcode`\@=11
\def\adv@nce{\global\advance\eqcount by 1}
\def\unadv@nce{\global\advance\eqcount by -1}
\def\nextnumber{\adv@nce}
%
%
\newif\iflines
\newif\ifm@resection
\def\onesec{\m@resectionfalse}
\def\moresec{\m@resectiontrue}
\moresec
\def\eq{\global\linesfalse\eq@}
\def\eqn{\global\linestrue&\eq@}
\def\nosubind@x{\global\subind@xfalse}
\def\newsubind@x{\ifsubind@x\unadv@nce\else\global\subind@xtrue\fi}
\newif\ifsubind@x
\def\eq@#1.#2.{\adv@nce
 \if\relax#2\relax
  \edef\loc@lnumber{\ifm@resection\number\secount.\fi
  \number\eqcount}
  \nosubind@x
 \else 
  \newsubind@x
  \edef\loc@lnumber{\ifm@resection\number\secount.\fi
  \number\eqcount#2}
 \fi
 \if\relax#1\relax
 \else 
  \expandafter\xdef\csname #1@\endcsname{{\rm(\loc@lnumber)}}
  \expandafter
  \gdef\csname #1\endcsname##1{\csname #1@\endcsname
  \ifcat##1a\relax\space
  \else
   \ifcat\noexpand##1\noexpand\relax\space
   \else
    \ifx##1$\space
    \else
     \if##1(\space
     \fi
    \fi
   \fi
  \fi##1}\relax
 \fi
 \eq@@{\loc@lnumber}}
\def\eq@@#1{\iflines \else \eqno\fi{\rm(#1)}}
\def\m@th{\mathsurround=0pt}
%
%
\def\display#1{\null\,\vcenter{\openup1\jot
\m@th
\ialign{\strut\hfil$\displaystyle{##}$\hfil\crcr#1\crcr}}
\,}
\newif\ifdt@p
\def\@lign{\tabskip=0pt\everycr={}}
\def\displ@y{\global\dt@ptrue \openup1 \jot \m@th
 \everycr{\noalign{\ifdt@p \global\dt@pfalse
  \vskip-\lineskiplimit \vskip\normallineskiplimit
  \else \penalty\interdisplaylinepenalty \fi}}}
%
%
\def\displayno#1{\displ@y \tabskip=\centering
 \halign to\displaywidth{\hfil$
\@lign\displaystyle{##}$\hfil\tabskip=\centering&
\hfil{$\@lign##$}\tabskip=0pt\crcr#1\crcr}}
%
%
\def\cite#1{{[#1]}}
\catcode`\@=\active
%
\hyphenation{Sebbar}
%
\magnification=\magstep1
\hsize= 6.75 true in
\vsize= 8.75 true in 
%
%
\def\RightHeadText{Modular Invariants and Generalized Halphen Systems}
\def\LeftHeadText{J. Harnad and J. McKay}
%
%
\leftline{solv-int/9902012 \hfill CRM-2597 (1999) \break} 
\bigskip \bigskip
\centerline{\titlefont Modular Invariants and }
\centerline{\titlefont Generalized Halphen
Systems\footnote{${}^{\dagger}$}{\eightpoint \hskip -8pt
Talk presented by J. Harnad at the SIDE III international meeting,
Sabaudia, May, 1998. Research supported in part by the Natural Sciences and
Engineering Research Council of Canada and the Fonds FCAR du Qu\'ebec.}}
\bigskip
\centerline{\authorfont J.~Harnad and J.~McKay}
\authoraddr
{Department of Mathematics and Statistics, Concordia University\\
7141 Sherbrooke W., Montr\'eal, Qu\'e., Canada H4B 1R6, {\rm \eightpoint
and} \\ 
Centre de recherches math\'ematiques, Universit\'e de Montr\'eal\\
C.~P.~6128, succ. centre ville, Montr\'eal, Qu\'e., Canada H3C 3J7\\
{\rm \eightpoint e-mail}: harnad\@ \hskip -3pt crm.umontreal.ca \quad
mckay\@cs.concordia.ca} 
\bigskip

\abstract{Generalized Halphen systems are solved in terms of
functions that uniformize genus zero Riemann surfaces, with automorphism
groups that are commensurable with the modular group.  Rational maps
relating these functions imply subgroup relations between their automorphism
groups and symmetrization relations between the associated
differential systems.} 
\bigskip \baselineskip 14 pt

\section 1. Introduction. Halphen Systems and Modular Invariants.

\Subtitle {1a. Darboux-Halphen Equations}
\smallskip
\nobreak

The Darboux--Halphen differential system: 
$$
\eqalign{
w_1' &=  w_1(w_2 + w_3) - w_2 w_3  \cr
w_2' &=  w_2(w_3 + w_1) - w_3 w_1    \cr
w_3' &=  w_3(w_1 + w_2) - w_1 w_2  } \eq Halphen.. 
$$
originally appeared in the work of Darboux \cite{Da} on orthogonal coordinate
systems. It was subsequently solved by Halphen \cite{Ha}, who  related it
to the hypergeometric equation of Legendre type
$$
\l(1-\l) {d^2 y \over d \l^2} + (1-2\l) {dy\over d\l} - {1\over 4} y = 0, 
\eq Legendrehypergeom..
$$
and also generalized it to a $3$--parameter family of systems
admitting a similar relation to the general hypergeometric equation
\cite{Ha, Br}. The general solution to \Halphen may be expressed \cite{Ha}
in terms of the {\it elliptic modular function}; that is, the square of the
elliptic modulus, viewed as a function of the ratio $\tau$ of the periods of
the Jacobi elliptic functions.  More recently, the system
\Halphen has found applications in mathematical physics in relation to
magnetic monopole dynamics
\cite{AH}, self--dual Einstein equations \cite{GP, Hi, To} and topological
field theory \cite{Du}.   

  To relate the Darboux--Halphen system to the hypergeometric equation
\Legendrehypergeom, we first form the ratio of two linearly independent
solutions of the latter
$$
\tau(\lambda):={y_1\over y_2},  \eq tauratio..
$$
and note that the inverse function $\l(\tau)$ satisfies the
Schwarzian equation \cite{GS, H}
$$
\{\l, \t\} + {\l^2 - \l + 1\over 2\l ^2(1 -\l)^2 }\l'^2 = 0, 
\eq Schwarzlambda..
$$
where the Schwarzian derivative is defined as
$$
\{f, \t\} := {f'''\over f'} -{3\over 2} \left({f''\over f'}\right)^2, 
\qquad  (f':={df\over d\t}). \eq.. 
$$
 In view of the invariance properties of the Schwarzian derivative,  the
general solution of \Schwarzlambda is obtained by composing a particular one
with the M\"obius (linear fractional)  transformations
$$
\t \ra {a\t  + b \over c \t + d} , 
\qquad \pmatrix{a & b \cr c & d } \in  SL(2, \bfC). \eq Mobius..
$$
The solutions to the Darboux-Halphen system are then given
\cite{Br, Ha} by setting 
$$
w_1 :=  {1\over 2}{d\over d\t}\ln {\l'\over \l},\qquad
w_2 :=  {1\over 2}{d\over d\t}\ln {\l'\over (\l-1)},\qquad
w_3 :=  {1\over 2}{d\over d\t}\ln {\l'\over \l(\l-1)}, \eq wdef..
$$
where $\lambda(\tau)$ is a solution of \Schwarzlambda. A particular
solution is provided \cite{H, GS} by the  ellliptic modular function 
$$
\lambda(\tau) = k^2(\tau), \eq..
$$
(with $\tau$ is interpreted as the ratio of two elliptic periods),
whose automorphism group is the principal congruence subgroup
$\Gamma(2)\ss\Gamma$ of the full modular group $\Gamma:= PSL(2,\bfZ)$.
 
An explicit representation of $\lambda(\tau)$ may be given in terms of 
null $\vartheta$-functions \cite{WW} 
$$
\l(\t) = {\vartheta_2^4(\tau) \over \vartheta_3^4(\tau)} 
= 1 - {\vartheta_4^4(\tau) \over \vartheta_3^4(\tau)}.   \eq..
$$
Substituting  this in \wdef  and using the differential identities
satisfied by the null theta functions leads to the explicit formulae
$$
w_1= 2 {d \over d\tau} \ln\th_4, \qquad w_2= 2 {d \over d \tau} \ln\th_2, 
\qquad w_3= 2 {d \over d\tau} \ln\th_3. \eq..
$$

\smallskip
\Subtitle {1b. Symmetrization under $\bfS_3=\Gamma/\Gamma(2)$. The Chazy
Equation}
\smallskip
\nobreak

The modular transformations
$$
\t \mt \ \tau+1, \  -{1\over \t}  \eq..
$$
do not leave $\l$ invariant, but generate the {\it group of anharmonic
ratios} \cite{H}
$$
\l \mt \l,\ {1\over \l}, \ 1-\l, \ {1 \over 1-\l}, 
\ {\l\over\l-1}, \ {\l-1\over \l}. \eq.. 
$$
The  symmetric invariant for this group is given by Klein's $J$--function
$$
J = {4(\l^2 -\l +1)^3\over 27 \l^2 (\l-1)^2} =
{ (\vartheta_2^8 + \vartheta_3^8 + \vartheta_4^8)^3\over 
54\vartheta_2^8 \vartheta_3^8 \vartheta_4^8}, \eq KleinJ..
$$
which has the full modular group $\Gamma$ as automorphism group, and satisfies
the Schwarzian equation
$$
\{J, \t\}+ { 36J^2 - 41J + 32\over 72 J^2(J-1)^2} J'^2 = 0. \eq SchwarzJ..
$$
In a similar manner to the above, this may be associated with the
hypergeometric equation
$$
J(1-J) {d^2y \over dJ^2} + \left({2\over 3} - {7\over 6} J\right) {dy\over dJ}
-{1\over 144} y = 0. \eq Jhypergeom.. 
$$

 In terms of the Halphen variables \wdef, it follows that the corresponding
elementary symmetric polynomials
$$
\s_1:= w_1 + w_2 +w_3, \qquad 
\s_2 := w_1 w_2 + w_2 w_3 + w_3 w_1,\qquad
\s_3 := w_1 w_2 w_3, \eq Halphensym.. 
$$
satisfy the symmetrized system
$$
\s'_1 = \s_2 \qquad
\s'_2 = 6\s_3 \qquad
\s'_3 =  4\s_1\s_3  - \s_2^2, \eq HalphenJ.. 
$$
which reduces to the Chazy  equation \cite{Ch} 
$$
W'''=2 W W'' - 3 W'^2 \eq Chazy..
$$
for
$$
W := 2\sigma_1 ={1\over 2}{d\over d\t} \ln{J'^6\over J^4(J-1)^3}. \eq..
$$

More generally, it is easy to see that, forming the ratio, as in \tauratio, of
a pair   $(y_1, y_2)$ of linearly independent solutions of the general
hypergeometric equation 
$$
f(1-f) {d^2y \over df^2} + (c - (a + b + 1) f) {d y \over df} - a b y =
0, 
\eq hypergeom..
$$
and {\it assuming} that the inverse function $f=f(\tau)$ is well--defined, 
this similarly provides solutions to the Schwarzian equation  
$$
\{f, \t\} +  {1\over 2}\left({1-\l^2\over f^2} + {1-\mu^2\over (f-1)^2} + 
{\l^2 +\mu^2 -\nu^2 -1 \over f(f-1)}\right) f'^2=0,  \eq Schwarzianeqhyper..
$$
where $(\l, \mu, \nu)$ are the relative Frobenius exponents for 
\hypergeom at $(0,1,\infty)$, given in terms of the parameters $(a,b,c)$ by 
$$
\l := 1-c, \qquad \mu := c-a-b, \qquad \nu := b-a . \eq.. 
$$
Introducing the general {\it Halphen-Brioschi variables} \cite{Br}
$$
W_1 := {1\over 2}{d\over d\t}\ln {f'\over f},\quad
W_2 := {1\over 2}{d\over d\t} \ln {f'\over (f-1)},\quad
W_3 := {1\over 2} {d\over d\t}\ln {f'\over f(f-1)}, \eq HalphenBrioschi..  
$$
these are similarly seen to satisfy the {\it general Halphen system}:
$$
\eqalign{
W_1' = &W_1( W_2+W_3) - W_2 W_3 + X(\lambda, \mu,\nu)\cr
W_2' = &W_2( W_3+W_1) - W_3 W_1 + X(\lambda, \mu,\nu) \cr
W_3' = &W_3( W_1+W_2) - W_1 W_2 + X(\lambda, \mu,\nu),} \eq generalHalphen..
$$
where
$$
\eqalign{X(\lambda, \mu,\nu) := &\mu^2 W_1^2 + \l^2 W_2^2 + \nu^2 W_3^2 +(\nu^2
- \l^2 -\mu^2)W_1 W_2 \cr &+(\l^2 - \mu^2 -\nu^2)W_3 W_1+(\mu^2 - \l^2
-\nu^2)W_2 W_3.}
\eq Halphenabc..  
$$

  This procedure, although formally identical to the two cases
treated above, can really only be viewed as providing
global solutions to such generalized systems if certain additional
conditions, relating to functional inversion and modularity of the resulting
functions, are satisfied. We therefore end  this introductory section by posing
the following questions:
\item{1.} 
 When is the functional inversion $\tau(f)\ra f(\tau)$ well-defined?\hfill 
\item {2.} 
 Are there other cases in which the resulting function $f(\tau)$ is a modular
function, with known properties, analogous to $\lambda$ and $J$?
\item {3.}  Are there any further generalizations of the above systems
admitting solutions in terms of modular functions? 

\noindent These questions will be addressed in the following sections.

\section 2. Modular solutions of general Halphen systems.

\Subtitle {2a. Replicable functions, Hauptmoduls and Schwarzian equations}
\smallskip
\nobreak

  A sufficient condition for the existence of a well-defined inverse
function $f(\tau)$ is that the (projectivized) monodromy group of the
hypergeometric equation in question be a Fuchsian group of the first type.
Essentially, this means that the possibly infinite multivaluedness
of the functions $\tau(f)$ defined by the ratio of two solutions can be
characterized by subdividing its image in the $\tau$--plane into fundamental
domains (each having a finite number of sides), which are permuted
amongst themselves by the action of the monodromy group, and into which each
branch of the function is mapped in a single--valued way. This
characterization may be applied not only to hypergeometric equations, but more
generally, to Fuchsian differential equations having an arbitrary number of
regular singular points, the number coinciding generally with the number of
vertices in a fundamental domain. In the case of hypergeometric equations,
there are three such singular points, at $(0,1,\infty)$, and the domains are
necessarily triangular. 

 In this section, a class of modular functions $f(\tau)$ is presented that
provide instances of globally defined inverses of the multivalued functions
$\tau(f)$ given by ratios of solutions of the hypergeometric equation
\hypergeom for certain particular parameter values $(a,b,c)$. By
``globally'', we here understand an open domain of definition with natural
boundary, in the interior of which the functions are holomorphic, and beyond
which they do not admit analytic continuation. As in the case of the
modular functions $\lambda$ and $J$, this domain in the first instance
consists of the upper half of the complex
$\tau$ plane, with the real axis as boundary, but upon application of the
M\"obius transformations \Mobius, it may become the interior of any disc.
In  subsequent sections,  examples of functions within the same general class
will be given that provide solutions to further generalizations of
Halphen systems.

  The particular class of functions to be considered here are the {\it
replicable functions} \cite{CN, FMN}, which arose originally in the context of
``modular moonshine''. These functions provide generalizations of the
$J$--function and are similarly defined in the upper half--plane by a
normalized $q$--series (\cite{CN, FMN}
$$
F(q) ={1\over q} + \sum_{n=1}^\infty a_n q^n, 
 \qquad q:= e^{2i\pi \t} \eq Fqseries..
$$
satisfying suitable invariance properties under generalized Hecke
operators. To relate these series to the functions obtained from solutions of
the hypergeometric equations, an affine transformation must be applied
$$
F(\tau) = a f(\tau) +b, \eq..
$$
with the constants $(a, b)$ chosen so that the values of $f$ at the
vertices of the fundamental domains are $(0,1,\infty)$. The main properties of
these functions that are of importance in what follows are: 
\item {1.} They are {\it uniformizing functions} for  genus
zero  Riemann surfaces $\grH/\grG_f$ formed by quotienting the upper
half--plane $\grH$ by the automorphism group $\grG_f$ of the function.
Such functions are referred to as {\it Hauptmoduls}.
\item {2.} The automorphism group $\grG_f$ is a subgroup of $PGL(2, \bfQ)$
that is {\it commensurable} with the modular group $\Gamma$; that is, the
intersection $\grG_f \cap \Gamma$ is of finite index in both.
\item {3.} In view of the form of the $q$--series \Fqseries, each such
function has a {\it cusp} at $\tau = i\infty$.
\item {4.}  The automorphism group $\grG_f$ in each case contains a subgroup
of type
$$
\G_0(N) :=\left\{\pmatrix {a & b \cr c & d } \in  SL(2,\bfZ), \ c\equiv 0 \
\mod N\right\}, \eq..
$$
with the smallest such $N$ referred to as the {\it level} of the function
(or group).
\item {5.} Finally, for the cases to be considered here, the coefficients
$a_n$ are all integers (although this is not part of the properties shared by 
all replicable functions).

\noindent A complete list of such replicable functions, together with many of
their properties, is provided in refs.~\cite{CN, FMN}.

   The fact that each such function is the generator of the field of
meromorphic functions on a genus zero Riemann surface implies that they
all satisfy a Schwarzian equation of the same general form as
\Schwarzianeqhyper
$$
\{f, \t\} + 2R(f) f'^2=0,  \eq Schwarzianeq..
$$
where $R(f)$ is {\it some} rational function. In the case where the
fundamental domains are triangular, with vertices mapping to $(0,1,\infty)$
in the $f$--plane, $R(f)$ will have the form appearing in
eq.~\Schwarzianeqhyper and the functions may be associated to solutions of a
corresponding hypergeometric equation.  The angles at the vertices of the
fundamental domain are determined in terms of the hypergeometric parameters as
$(\l\pi,\mu\pi ,\nu \pi)$. In general,
$R(f)$ could have any number of poles at arbitrary locations, but  in the case
of the replicable functions arising in \cite{CN, FMN}, this number never
exceeds $25$. The remainder of this section concerns only the triangular case;
in the following sections, cases with higher numbers of vertices and the
corresponding generalized Halphen systems will be discussed.

\Subtitle {2b. Triangular replicable functions}
\smallskip
\nobreak

 Table 1 below, which is taken from ref.~\cite{HM}, lists all cases, up to
equivalence under M\"obius transformations \Mobius and affine transformations
of $f$, of replicable triangular functions with integer $a_n$'s. These are
the modular functions whose automorphism groups are the arithmetic
triangular groups of noncompact type classified in \cite{Ta}.

 The first column identifies the groups according to the notation of \cite{CN,
FMN}, with the integer indicating the level. The second and third columns give
the associated hypergeometric parameters and angles at the vertices
$(0,1,\infty)$, respectively. The fourth column gives the generators
$(\rho_0$ , $\rho_1)$ of the automorphism group fixing a pair of finite vertices
mapping to the points $0$ and $1$ in the $f$--plane. The third generator,
stabilizing the point $i \infty$, is
$$
\rho_\infty\ =\smatrix[ 1\& 1\\ 0\& 1], \eq..
$$
and the three generators together satisfy the consistency relation
$$ 
\rho_\infty \rho_1 \rho_0 = \bfI. \eq..
$$

The fifth column in the table indicates the relation between the
function $F$ normalized as in \Fqseries and $f$ normalized as in the
hypergeometric equation, so that the vertices are mapped to $(0,1,\infty)$.
The last column gives explicit expressions for the modular function $f$ in
terms of null theta functions or the Dedekind eta function $\eta(\tau)$.
In this table, the case $1A$ is just the $J$ function, which therefore
determines solutions to the symmetrized Halphen system \HalphenJ, while the
case 4C is essentially the function $\l$ (more precisely, composed with the
transformation $\tau \ra 2\tau$, $\l \ra 1/\l$ as indicated in the sixth
column), and hence determines solutions to the original Darboux--Halphen
system \Halphen. The other cases provide seven further examples of globally
defined modular solutions to the general Halphen system \generalHalphen.

\vskip 12pt
\centerline{\bf{Table 1.  Triangular Replicable Functions}}
\nobreak
\vskip 12pt 
\centerline{\hskip -5pt
\vbox{\tabskip=0pt \offinterlineskip
\def\tablerule{\noalign{\hrule}}
\halign to 455pt{\strut#& \vrule#\tabskip=.5em plus2em&
 \hfil#\hfil & \vrule # &\hfil #\hfil & \vrule # &
 \hfil#\hfil & \vrule# & \hfil#\hfil & \vrule# &
 \hfil#\hfil & \vrule# & \hfil#\hfil & \vrule#
\tabskip=0pt\cr\tablerule
&& Name && $(a,b,c)$ && $(\lambda, \mu, \nu)$ && $\rho_0 \qquad \quad \rho_1$ 
&& $ F $ && $f(\tau)$   &\cr \tablerule
&& $\matrix{1A \cr \sim \G}$ && $({1\over 12},{1\over 12},{2\over 3})$ 
&& $({1\over 3},{1\over 2},0)$  &&
$\smatrix[ 0 \&-1 \\ 1 \& -1] $ \ $\smatrix[ 0 \&-1 \\ 1 \& \phantom{-}0] $
&& $ 1728 f -744$  && $J = {(\vartheta_2^8 + \vartheta_3^8 + 
\vartheta_4^8)^3\over  54 \vartheta_2^8 \vartheta_3^8 \vartheta_4^8}$
&\cr \tablerule
&& $2A$ && $({1\over 8}, {1\over 8}, {3\over 4})$ 
&& $({1\over 4}, {1\over 2}, 0)$ &&
$\smatrix[ 0 \&-1 \\ 2 \& -2] $ \ $\smatrix[ 0 \&-1 \\ 2 \& \phantom{-}0] $
 && $256 f -104$ &&  
${ \left(\vartheta_3^4 +\vartheta_4^4\right)^4 \over
16  \vartheta_2^8 \vartheta_3^4 \vartheta_4^4}$
&\cr\tablerule 
&& $3A$ && $({1\over 6}, {1\over 6}, {5\over 6})$ && 
$({1\over 6}, {1\over 2}, 0)$  && 
$\matrix{\cr\smatrix[ 0 \&-1 \\ 3 \& -3] $ \ $\smatrix[ 0 \&-1 \\ 3 \&
\phantom{-}0]\cr\phantom{m} }$
 && $108 f -42 $  &&  ${\left(\eta^{12}(\tau) + 27 \eta^{12}(3\tau)\right)^2
\over 108 \eta^{12}(\tau)\eta^{12}(3\tau)}$ 
&\cr\tablerule
&& $\matrix{2B \cr \sim\G_0(2)}$ && $({1\over 4}, {1\over 4}, {1\over 2})$ && 
$({1\over 2}, 0, 0)$  && 
$\matrix{\cr\cr\smatrix[ 1 \&-1 \\ 2 \& -1] $ \ $\smatrix[ -1 \& \phantom{-}0 
\\ \phantom{-}2 \& -1]\cr\cr\phantom{m}}$
 && $64 f -40  $  
&&  $\matrix{1 + {1\over 64}\big({\eta(\tau)\over
\eta(2\tau)}\big)^{\scriptscriptstyle 24}\cr = {\big(\vartheta_3^4(\tau) +
\vartheta_4^4(\tau)\big)^2\over
\vartheta_2^8(\tau)}} $
&\cr\tablerule
&& $\matrix{3B \cr \sim \G_0(3)}$ && $({1\over 3}, {1\over 3}, {2\over 3})$ && 
$({1\over 3}, 0, 0)$  && 
$\smatrix[ 1 \&-1 \\ 3 \& -2] $ \ $\smatrix[ -1 \& \phantom{-}0 \\ 
\phantom{-}3 \& -1]$
 && $ 27 f -15  $  &&  $1 + {1\over 27}\big({\eta(\tau)\over
\eta(3\tau)}\big)^{\scriptscriptstyle 12}$ 
&\cr\tablerule
&& $\matrix{4C^*\cr \sim \G_0(4)}$ && $({1\over 2}, {1\over 2}, 1)$ && 
$(0, 0, 0)$  && 
$\matrix{\cr\smatrix[ 1 \&-1 \\ 4 \& -3] $ \ $\smatrix[ -1 \& \phantom{-}0 \\ 
\phantom{-}4 \& -1 ]\cr\cr}$
 && $16 f - 8 $  &&  $\matrix{{1\over \l(2\tau)}
 = {\vartheta_3^4(2\tau) \over \vartheta_2^4(2\tau)}\cr 
=1 +{1\over 16}\big({\eta(\tau)\over\eta(4\tau)}\big)^{\scriptscriptstyle 8}
\cr }$ &\cr\tablerule
&& $2a$ && $({1\over 6}, {1\over 6},  {2\over 3})$ && 
$({1\over 3}, {1\over 3}, 0)$  && 
$\matrix{\cr\smatrix[ 2 \&-3 \\ 4 \& -4] $ \ $\smatrix[ 0 \&-1 \\ 4 \&
-2]\cr\phantom{m}}$
 && $24\sqrt{3}i(2 f - 1) $  &&  
${\sqrt{3}i\left( e^{\pi i/3}\th_3^4(2\tau) -\th_2^4(2\tau)\right)^3\over
9\th_2^4(2\tau)\th_3^4(2\tau)\th_4^4(2\tau)}$
&\cr\tablerule
&& $4a$ && $({1\over 4}, {1\over 4}, {3\over 4})$ && 
$({1\over 4}, {1\over 4}, 0)$  && 
$\matrix{\cr\smatrix[ 4 \&-5\\ 8 \& -8] $ \ $\smatrix[ 0 \&-1 \\ 8 \&
-4]\cr\phantom{m}}$
 && $-16i(2 f - 1)$  && $-{i\left(\th_3^2(2\tau) +
i \th_4^2(2\tau)\right)^4\over 8\th_2^4(2\tau)\th_3^2(2\tau)\th_4^2(2\tau)}$
&\cr\tablerule
&& $6a$ && $({1\over 3}, {1\over 3}, {5\over 6})$ && 
$({1\over 6}, {1\over 6}, 0)$  && 
$\matrix{\cr\smatrix[ 6 \&-7 \\ 12 \& -12] $ \ $\smatrix[ 0 \&-1 \\ 12 \&
-6]\cr\phantom{m}}$
 && $ 6\sqrt{3}i( 2f - 1)$  &&  
$-{\sqrt{3}i \left(\eta^6(2\tau) + 3\sqrt{3}i\eta^6(6\tau)\right)^2\over
36\eta^6(2\tau)\eta^6(6\tau)}$
&\cr\tablerule 
\hfil \cr}}}
\vskip 10pt

  Although these cases are distinct, it is important to note that, just as the
functions $\lambda$ and $J$ are related by the rational map defined in
\KleinJ, so all these replicable functions are linked by algebraic
relations, some of which may similarly be expressed in terms of explicit
rational maps. This implies that the general Halphen--Brioschi variables for
the various cases are also algebraically related, just as in the
Halphen-Chazy case, where the ones for the $J$ case are essentially
the elementary symmetric invariants of those for the $\lambda$ case. A further
illustrative example of such algebraic relations is given in the next
subsection; the full set of rational maps relating various triangular
cases are given in \cite{HM}, to which the reader is referred for details.

\Subtitle {2c. Rational transformations}
\smallskip
\nobreak
If $f(\tau)$ satisfies a Schwarzian equation of the form \Schwarzianeqhyper
(where $R(f)$ is any rational function) and there is a map
$$
f = Q(g), \eq rationalmap..
$$
relating it to another function $g(\tau)$,
where $Q(g)$ satisfies the Schwarzian equation
$$
\{Q, g\} + 2 R(Q(g)) Q'^2 = 2\tilde{R}(g), \eq..
$$
for some function $\tilde{R}(g)$,
then $g$  satisfies the transformed Schwarzian equation
$$
\{ g, \tau\} + 2 \tilde{R}(g) g'^2 = 0.\eq..
$$
Also, if $y(f)$ satisfies the second order linear equation
$$
{d^2 y \over df^2} + R(f)y = 0, \eq fFuchsianeq..
$$
then $\tilde{y}(g):=(Q')^{-{1\over 2}}y(Q(g))$ satisfies the transformed
equation:
$$
{d^2 \tilde{y} \over dg^2} + \tilde{R}(g) \tilde{y} = 0.\eq gFuchsianeq..
$$

Similar transformations may be applied to second order equations in which 
first order derivative terms are also present. For the case of hypergeometric 
equations, such transformations, with $Q(g)$ rational functions of degree up
to four,  were studied by Goursat \cite{Go} in his thesis. They imply
corresponding {\it symmetrizations} of the associated general Halphen systems,
such as the ones relating \Halphen and \HalphenJ. These are discussed in detail
in ref.~\cite{HM}; here, we present just one illustrative example taken from
\cite{HM}.

\noindent Example 2.1. $ 3B \mt 1A. $ \hfill

\nobreak
In this case the relevant identity  relating the associated hypergeometric
functions is 
$$
F\left({1\over 12}, {1\over 12}; {2\over 3}; {x (x+8)^3\over
64(x-1)^3}\right) 
= (1-x)^{{1\over 4}}F\left({1\over 3}, {1\over 3}; {2\over 3}; x\right).
\eq..  
$$
The corresponding rational map $Q(g)$ is therefore defined by
$$
 f= {g(g+8)^3\over  64(g-1)^3}, \eq..
$$
where $f$ denotes the Hauptmodul of case $1A$ (i.e., the modular function $J$),
and $g$ is the Hauptmodul of case $3B$.  The relationship between the
corresponding automorphism groups is given by a symmetrization
quotient which may also be expressed as a quotient of finite groups: 
$$
\grG_f/\grG_g= \G/ \G_0(3) =\bfS_{1A}^{9B}/\bfS_{3B}^{9B}=\bfA_4/\bfZ_3
\eq..
$$
(The subgroup $\grG_g$ for this case is not normal in $\grG_f$,
so the corresponding field extension is non-Galois. The
expressions $\bfS_{1A}^{9B}$, $\bfS_{3B}^{9B}$ denote the finite groups
obtained by quotienting the respective automorphism group of $1A$ and $3B$
by the largest subgroup that is normal in both, which in this case also
corresponds to a replicable function, denoted by $9B$, appearing in Table 2
below. ) We may associate the following  polynomial invariants to this
symmetrization: 
$$
\eqalign{
\S_1 := &3w_1 + 2w_2 + w_3, \cr \S_2 := &(w_1 - w_3)(9w_1 - 8w_2 - w_3), 
\cr \S_3 :=&
 (w_1 - w_3)(27w_1^2 - 36 w_1 w_2 + 8w_2^2 - 18 w_1 w_3 + 2 w_2 w_3 -
w_3^2),} \eq..
$$
where $(w_1,w_2,w_3)$ are the Halphen--Brioschi variables for the
case $3B$, in terms of which the corresponding variables for the
case $1A$, denoted $(W_1, W_2, W_3)$ are determined by
$$
\eqalign{
3W_1 + 2W_2  + W_3 & = \phantom{-}\S_1  \cr
W_1  - W_3  & = -{\S_2^2 \over \S_3}  \cr
W_2 - W_3  & = -{\S_3\over \S_2}. }  \eq..
$$

\section 3.  Generalized Halphen Systems.

\Subtitle {3a. Fuchsian equations, monodromy and automorphism groups}
\smallskip
\nobreak

Up to projective equivalence (i.e., multiplication of solutions by a
common function that does not alter the location of the singular
points), second order Fuchsian equations may be expressed in the form
\fFuchsianeq, where $R(f)$ is a rational function of the form 
$$
R(f) ={N(f)\over (D(f))^2}, \qquad
D(f) = \prod_{i=1}^n (f-a_i),  \eq RNumDenom..
$$
and the numerator is a polynomial  $N(f)$ of degree $\leq 2n - 2$. The
ratio of two linearly independent solutions 
$$
\tau(f):={y_1\over y_2} \eq tauratio..
$$
then  satisfies the Schwarzian differential equation \cite{GS, H} 
$$
\{\t, f\} = 2 R(f),   \eq tauSchwarzeq..
$$
and the inverse function $f=f(\tau)$ (if well defined) satisfies
eq.~\Schwarzianeq with $R(f)$ given by \RNumDenom.
The image of the monodromy representation for eq.~\fFuchsianeq
determines a subgroup $\grG_f \ss PGL(2, \bfC)$ that acts on the
ratio $\tau$ of solutions by linear fractional transformations \Mobius
leaving the inverse function $f(\tau)$ invariant. 

\smallskip
\Subtitle {3b.  Constrained Dynamical Systems and Flows on the $SL(2,\bfC)$ 
Manifold}
\smallskip
\nobreak

Following Ohyama \cite{Oh}, we may associate to any such Fuchsian system a
dynamical system in $n+1$ variables, subject to $n-2$ independent
quadratic constraints. Define the following $n+1$ variables, which
serve to generalize the Halphen--Brioschi variables,
$$
X_0 := {1\over 2} {d\over d \tau}\ln f', \quad 
X_i := {1\over 2}{d \over d \tau} \ln {f'\over (f- a_i)^2}, \quad i=1, 
\dots n.  \eq Ohyamavars..
$$
Equivalently, we may use the linear combinations
$$
u:= X_0 = {1\over 2} {f'' \over f'}, \quad v_i :={1\over 2}( X_0 - X_i) =
{1\over 2}{f'\over f- a_i}. \eq..
$$
These satisfy the set of quadratic constraints 
$$
(a_i - a_j) v_i v_j + (a_j - a_k) v_j v_k
+ (a_k - a_i) v_k v_i  = 0, \qquad 1\leq i, j,k \leq n \eq quadconstr..     
$$
(of which $n-2$ are independent), and the differential equations:
$$
\eqalignno{
v'_i&= -2 v_i^2 + 2 u v_i, \qquad i=1,\dots n,   \eqn veq.a.  \cr
u' &= u^2 - \sum_{i,j=1}^n r_{ij} v_i v_j,   \eqn ueq.b.} 
$$
where the coefficients $r_{ij}$ defining the quadratic form 
$$
r(v):= \sum_{i,j=1}^n r_{ij} v_i v_j
$$ 
appearing in \ueq  are obtained by expressing $R(f)$ in the form
$$
R(f) = {1\over 4}\sum_{i,j=1}^n {r_{ij}\over (f-a_i)(f- a_j)}. \eq..
$$
(This leaves a residual ambiguity in their definition, which just
consists of adding any linear combination of the vanishing quadratic forms
appearing in  \quadconstr. The definition may be made unique by
fixing an ordering for the singular points, and requiring that all
coefficients $r_{ij}$ vanish, except when $i= j, j\pm 1$.) The quadratic form
$r(v)$ encodes all the relevant information about the monodromy of 
the associated Fuchsian system and, in the cases where the associated group is
modular,  about the geometry of the fundamental domains. In particular, the
angles at the vertices are $\{\a_i \pi\}_{i=1, n}$ where
$$
r_{ii} = 1 -\a_i^2.  \eq..
$$

  Equivalently, these systems may be viewed as unconstrained 
dynamical systems on the $SL(2,\bfC)$ group manifold. To see this, let
$$
g(\tau) := \pmatrix { A & B \cr C & D} \in SL(2,\bfC)   \eq groupel..
$$
denote an integral curve in $SL(2,\bfC)$ for the equation
$$
g' = \pmatrix{ 0 & \g \cr -1  & 0 } g,  \eq..
$$
where
$$
\g := -{1\over 4C^2} \sum_{i,j =1}^n {r_{ij}\over (Ca_i + D)(C a_j + D)}
= -{1\over C^2} R\left(-{D \over C}\right). 
\eq..
$$
Defining $f(\tau)$ as
$$
f := - {D \over C},  \eq..
$$
it follows that this satisfies \Schwarzianeq, and that
$$
u := {A\over C}, \qquad v_{a_i} := {1\over 2C(Ca_i +D)}, \qquad i=1, \dots n,
\eq..
$$
satisfy the system \veq, \ueq and the constraints \quadconstr .
Ohyama's variables \Ohyamavars are recovered by applying $g$ as a linear
fractional transformation to $\{\infty, a_1, \dots a_n\}$.
$$
X_0 = {A\over C}, \qquad X_i = {A a_i + B \over C a_i + D},
 \qquad i=1, \dots n.  \eq..
$$

\Subtitle {3c.  Examples  with Four Vertices}
\smallskip
\nobreak

 Table 2 below, which is also taken from \cite{HM}, lists all cases of
replicable functions having integer $q$--series coefficients, 
whose fundamental domains have four vertices, and which are related by a
rational map of degree $\leq 4$ to one of the triangular cases.
These functions,  composed with the M\"obius transformation \Mobius, therefore
provide the general solutions to generalized Halphen systems of
the type \quadconstr--\ueq,  with $n=3$. 

\bigskip
\centerline{\bf{Table 2.  Four Vertex Replicable Functions}}
\nobreak
\centerline{{\smaller (admitting a rational map of degree 
${\scriptstyle \leq 4}$ to a triangular replicable function)}}
\nobreak\medskip \smallskip 
\centerline{
\vbox{\tabskip=0pt \offinterlineskip
\def\tablerule{\noalign{\hrule}}
\halign to430pt{\strut#& \vrule#\tabskip=.5em plus1em&
 \hfil#\hfil & \vrule # &\hfil #\hfil & \vrule # &
 \hfil#\hfil & \vrule# & \hfil#\hfil & \vrule# &
 \hfil#\hfil & \vrule# & \hfil#\hfil & \vrule#
\tabskip=0pt\cr\tablerule
&& Name && $(a_1,a_2,a_3)$ && $\matrix{\rho_1 \cr \rho_2 \cr \rho_3}$ 
&& $\displaystyle{ \sum_{i,j=1}^3 r_{ij}v_i v_j} $ && $ F $ &&$f(\tau)$  
&\cr \tablerule
&& $\matrix{6C}$ && $(-3,0,1)$ && $\matrix{\smatrix[3 \& -2\\ 6 \& -3]\cr 
\smatrix[3 \& -1 \\ 12 \& -3]\cr 
\smatrix[-1 \& \phantom{-}0\\ \phantom{-}6 \& -1]}$  
&& $\matrix{{3\over 4}v_1^2 + {3\over 4} v_2^2 + v_3^2\cr
 - {1\over 2} v_2 v_3 - v_1 v_3}$ 
&& $4f + 2$  && $1 + {1\over 4} {\eta^6(\t) \eta^6(3\t) \over \eta^6(2\t)
\eta^6(6\t)}$ 
&\cr\tablerule
&& $\matrix{6D}$ && $\matrix{(\b,\bar{\b},1) \cr
\b:=-{3\over 4} + \sqrt{2}i}$ 
&&
$\matrix{ \smatrix[4 \& -3 \\ 6 \& -4]\cr
\smatrix[2 \& -1\\ 6 \& -2]\cr \smatrix[-1 \& \phantom{-}0\\
\phantom{-}6 \& -1]}$  && $\matrix{{3\over 4}v_1^2 + {3\over 4} v_2^2 + v_3^2
\cr
 + {131\over 162} v_1 v_2 \cr - {28 -16\sqrt{2}i\over 81}v_1 v_3 \cr
- {28 +16\sqrt{2}i\over 81}v_2 v_3}$ 
&& $4f$  && $1 + {1\over 4} {\eta^4(\t) \eta^4(2\t) \over \eta^4(3\t)
\eta^4(6\t)}$ 
&\cr\tablerule
&& $\matrix{6E\cr \sim \G_0(6)}$ && $(-{1\over 8},0,1)$
 && $\matrix{\smatrix[5 \& -3\\ 12 \& -7]\cr 
\smatrix[5 \& -2 \\ 18 \& -7]\cr 
\smatrix[-1 \& \phantom{-}0\\ \phantom{-} 6 \& -1]}$  
&& $\matrix{v_1^2 + v_2^2 + v_3^2\cr
 - {10\over 9} v_2 v_3 -{8\over 9} v_1 v_3}$ 
&& $8f -3$  && $1 + {1\over 8} {\eta^5(\t) \eta(3\t) \over \eta (2\t)
\eta^5(6\t)}$ 
&\cr\tablerule
&& $\matrix{6c}$ && $(-1,1,0)$ && $\matrix{\smatrix[8 \& -7 \\ 12 \& -10] \cr
\smatrix[2 \& -1\\ 12 \& -4] \cr  \smatrix[-1 \& \phantom{-}0\\ 12 \& -1]}$ &&
$\matrix{{8\over 9}v_1^2 + {8\over 9} v_2^2 + v_3^2\cr
 +{16\over 9} v_1 v_2}$ 
&& $i3\sqrt{3}f$  && $-{i\over 3\sqrt{3}} {\eta^6(2\t) \over \eta^6(6\t)}$ 
&\cr\tablerule
&& $\matrix{8E\cr \sim \G_0(8)}$ && $(-1,0,1)$ 
&& $\matrix{\smatrix[3 \& -2\\ 8 \& -5]\cr 
\smatrix[3 \& -1 \\ 16 \& -5]\cr 
\smatrix[-1 \& \phantom{-}0 \\ \phantom{-}8 \& -1]}$  
&& $\matrix{v_1^2 +  v_2^2 + v_3^2\cr - 2v_1 v_3 }$ && $4f $   
&& $\matrix{1 + {1\over 4} {\eta^4(\t) \eta^2(4\t) \over
\eta^2(2\t) \eta^4(8\t)}\cr
={\th_3^2(2\t) + \th_4^2(2\t) \over \th_3^2(2\t) - \th_4^2(2\t)}}$ 
 &\cr\tablerule
&& $\matrix{9B\cr \sim \G_0(9)}$ 
&& $\matrix{(\o,\bar{\o},1)\cr \o:= e^{2\pi i\over 3}}$ &&
$\matrix{ \smatrix[5 \& -4 \\ 9 \& -7]\cr \smatrix[2 \& -1\\ 9 \& -4]\cr 
\smatrix[-1 \& \phantom{-}0\\ \phantom{-}9 \&-1]}$  
&& $\matrix{v_1^2 +  v_2^2 + v_3^2\cr
 - v_1 v_2 \cr - (1-\o)v_1 v_3 \cr - (1-\bar{\o})v_2 v_3}$ 
&& $3f$  && $1 + {1\over 3} {\eta^3(\t) \over \eta^3(9\t)}$ 
&\cr\tablerule
\hfil\cr}}}
\medskip 

The first column again gives the label of the associated automorphism group 
$\grG_f$ for each function $f(\tau)$ in the notation of \cite{CN, FMN}, while
the second specifies the location of the finite singular points $(a_1,a_2,a_3)$
of the associated Fuchsian equation \fFuchsianeq. The third column gives the
automorphism group generators stabilizing three finite vertices mapping to
$(a_1, a_2, a_3)$, and the fourth gives the quadratic form $r(v)$ appearing in
eq.~\ueq. The fifth column lists the affine relations connecting the 
normalized $q$--series of the form \Fqseries with the functions $f(\tau)$
mapping the vertices to the singular  points $(a_1,a_2,a_3)$  of \fFuchsianeq,
and the last column gives explicit expressions for $f(\tau)$ in terms of the
Dedekind $\eta$--function.   

There are also algebraic relations connecting these dynamical systems with
triangular cases, following from the existence of rational maps satisfying
\rationalmap--\gFuchsianeq. The following example is an illustration of such
rational maps and the implied relations between the generalized
Halphen--Brioschi variables. (Further details and a list of the other cases 
may be found in \cite{HM}.)

\noindent Example 3.1. $8E \mt 4C$. \hfill 
 
The rational map \rationalmap relating these two cases is given by
$$
 f = {(g + 1)^2\over 4g}, \eq.. 
$$
where $f$ denotes the Hauptmodul for the case $4C$ and $g$ the one for $8E$.
The automorphism group for $4C$ is $\Gamma_0(4)$, and that for $8E$ is
$\Gamma_0(8)$, a normal subgroup, and therefore in this case  the function field
generated by $g$ is a Galois extension of the one generated by $f$. The map is
of degree two and the quotient group characterizing the symmetrization is 
$$
\bfS_{4C}^{8E}= \G_0(4) / \G_0(8) =\bfZ_2, \eq..
$$ 
whose action is generated by 
$$
\t \mt  {-\tau\over 4\tau -1}, \qquad g \mt {1\over g}. \eq..
$$
The effect of this on the generalized Halphen--Brioschi variables is 
$$
(u, v_{-1}, v_0, v_1 ) \mt (u-2v_0,v_1 - v_0, -v_0, v_1 - v_0), \eq..  
$$
and the associated polynomial invariants are
$$
\S_1 := u + v_{-1} - v_0 -v_1, \quad \S_1':=4v_1-2v_0 ,
\quad \S_2 := 4v_0^2.  \eq..
$$
The Halphen--Brioschi variables for the $4C$ case are determined in terms of
these by
$$
W_1  = \S_1,  \quad
W_1  - W_3   = {\S_2 \over \S_1'}, \quad
W_2 - W_3   = \S_1'.  \eq..
$$

\Subtitle {3d. Two examples with $26$ vertices}
\smallskip
\nobreak

In the list of replicable functions with integer $q$--series coefficients,
\cite{CN, FMN}, there are three cases, denoted $72e$, $96a$ and
$144^{\scriptscriptstyle\sim}e$,  for which the fundamental domains have $26$
vertices, the maximal number occurring. We give here the corresponding
generalized Halphen--Brioschi variables, and the quadratic forms determining
the associated constrained  dynamical system for the cases $72e$ and $96a$.
(The case $144^{\scriptscriptstyle\sim}e$ is related to $72e$ by 
multiplication of $f$ by $e^{i\pi\over 12}$.)

\noindent Example 3.2. $72e$. \
The modular function for this case may be expressed as follows in terms of
the Dedekind $\eta$--function 
$$
f = {\eta(24\tau) \eta (36 \tau) \over \eta(12\tau) \eta(72\tau)}. \eq..
$$
The fundamental domain has $25$ finite vertices, and the rational function
$R(f)$ to whose poles these are mapped is 
$$
R(f) = {1\over 4f^2}\left(1 + 
{2^7 3^3f^{12}(f^{12} +1)^2\over
(f^{24}-34f^{12} +1)^2}\right).  \eq..
$$
The poles are located at the origin, and at the vertices of two
regular dodecagons centered at the origin, at radial distances
$(\sqrt{2}\pm1)^{1\over 3}$, forming angles that are  multiples of
$\pi/12$ with the axes:
$$
a_0 := 0, \quad
a_m := e^{(m-1)\pi i\over 6}(\sqrt{2}-1)^{1\over 3}, \quad
a_{12+m} := e^{(m-1)\pi i\over 6}(\sqrt{2}+1)^{1\over 3}, \qquad m=1, \dots
12.  \eq..
$$
The quadratic form $r(v)$ defining the dynamical system in this case is
$$
\sum_{i,j=1}^nr_{ij}v_i v_j = v_0^2 + {3\over 4} \sum_{m=1}^{24} v_m^2 
- {3\over 8} \sum_{m=1}^{11}(1-e^{m\pi i\over 6})
\big((2 -\sqrt{2} )v_m v_{m+1} + (2 +\sqrt{2} )v_{12+m} v_{13+m} \big).
\eq quadformste..
$$

\noindent Example 3.3. $96a$.\ For this case, the modular function  may be
expressed as 
$$
f = {\eta^2(48\tau)\over \eta(24\tau) \eta(96\tau)}, \eq..
$$
and the rational function $R(f)$ is 
$$
R(f) = {1\over 4f^2}\left( 1 + 
{2^{10}3^3f^{24}\over(f^{24}-2^6)^2}\right).  \eq..
$$
The poles are therefore located at the origin, and at the vertices of a
regular $24$--gon centered at the origin, at radial distance
$2^{1\over 4}$, forming angles that are  multiples of $\pi/24$ with the
axes:
$$
a_0 := 0, \quad
a_m := e^{(m-1)\pi i\over 12}2^{1\over 4},  \qquad m=1,
\dots 24.  \eq..
$$
The quadratic form $r(v)$ defining the dynamical system for this case is
$$
\sum_{i,j=1}^nr_{ij}v_i v_j = v_0^2 + {3\over 4} \sum_{m=1}^{24} v_m^2 
- {3\over4} \sum_{m=1}^{23}(1-e^{m\pi i\over 12})v_m v_{m+1}.
\eq..
$$

\bigskip \bigskip
  \centerline{\bf References}  
\medskip
 {\eightpoint
\item{\bf [AH]} Atiyah, M.F., and Hitchin, N.J., {\it The Geometry and 
Dynamics of Magnetic Monopoles}, Princeton University Press, Princeton (1988).
 \item{\bf[Br]} Brioschi, M.,  ``Sur un syst\`eme d'\'equations 
diff\'erentielles'', {\it C.~R.~Acad.~Sci.~Paris\/} {\bf 92}, 1389--1393
(1881).
 \item{\bf[Ch]} Chazy, J., ``Sur les \'equations diff\'erentielles dont 
l'int\'egrale g\'en\'erale poss\`ede une coupure essentielle mobile'',
{\it C.~R.~Acad.~Sc.~Paris}, {\bf 150}, 456--458 (1910). 
 \item{\bf[CN]} Conway, J. and Norton, S.~P., ``Monstrous moonshine'', 
{\it Bull.~Lond.~Math.~Soc.} {\bf 11}, 308--339 (1979).
\item {\bf[Da]} Darboux, G. {\it Le\c cons sur les syst\`emes othogonaux} (2nd
ed.). Gauthiers-Villars, Paris (1910).
\item{\bf [Du]} Dubrovin, B.A., ``Geometry of $2D$ topological field
theories'', Lecture Notes in Math. {\bf 1620}, Springer-Verlag, Berlin,
Heidelberg,  New York (1996). 
\item{\bf [FMN]} Ford, D., McKay, J.,  and Norton, S., ``More on replicable 
functions'' {\it Comm.~in Algebra} {\bf 22}, 5175--5193 (1994).
\item{\bf [GP]} Gibbons, G.W., and Pope, C.N., ``The Positive Action
Conjecture and Asymptotically Euclidean Metrics in Quantum Gravity'',  {\it
Commun.~Math.~Phys.} {\bf 66}, 267--290 (1979). 
\item{\bf [Go]} Goursat, E. ``Sur L'\'Equation diff\'erentielle lin\'eaire 
qui admet pour int\'egrale la s\'erie hyperg\'eom\'etrique'', {\it
Ann.~Sci.~de l'\'Ecole Normale Sup\'erieure}, {\bf X} suppl., 1--142 (1881).
\item {\bf [GS]} Gerretson, J.,  Sansone, G., {\it Lectures on the Theory of
Functions of  a Complex Variable. II. Geometric Theory}. Walters--Noordhoff,
Gr\"oningen (1969).
\item{\bf [H]}  Hille, Einar
{\it Ordinary Differential equations in the Complex Domain} ,  (Dover, New
York  1976), Sec. 7.3, Ch.~10; {\it Analytic Function Theory}, Vol. II. 
(Chelsea, New York  1973).
\item{\bf [Ha]} Halphen, G.-H., ``Sur des fonctions qui proviennent de 
 l'\'equation de Gauss'',  {\it C. R. Acad. Sci. Paris} {\bf 92},
856--858 (1881); ``Sur un syst\`eme d'\'equations   diff\'erentielles'',
{\it ibid.} {\bf 92}, 1101--1103 (1881);
``Sur certains syst\`emes d'\'equations  diff\'erentielles'',
{\it ibid.} {\bf 92}, 1404--1406 (1881).
\item{\bf [Hi]} Hitchin, N, ``Twistor Spaces, Einstein metrics and
isomondromic deformations'', {\it J.~Diff.~Geom.} {\bf 42}, 30--112 (1995).
\item{\bf [HM]} Harnad, J. and McKay, J., ``Modular Solutions to Equations
of Generalized Halphen Type'', preprint CRM-2536 (1998),  solv-int/98054006
\item{\bf [Oh]} Ohyama, Yousuke, ``Systems of nonlinear differential 
equations related to second order linear equations'', {\it Osaka J.~Math.}
{\bf 33}, 927--949 (1996);``Differential equations for modular forms with 
level three'', Osaka Univ. ~preprint (1997).
\item{\bf [Ta]} Takeuchi, K., ``Arithmetic triangle groups'', {\it
J.~Math.~Soc.~Japan}  {\bf 29}, 91--106 (1977).
\item{\bf [To]} Tod, K.P., ``Self--dual Einstein metrics from the Painlev\'e VI
equation'', {\it Phys.~Lett.} {\bf A190}, 3--4 (1994).
\item{\bf [WW]} Whittaker, E.T., and Watson, G.N., {\it A Course in Modern
Analysis}, Chapt.~21,  Cambridge University Press, 4th ed., London, N.Y.
(1969).

}
\vfill \eject

\end